\documentclass{article}
 \usepackage{authblk} 
 \usepackage{blindtext} 
 \usepackage{verbatim}
 \usepackage{graphicx}
 \usepackage{color}
 \usepackage[svgnames]{xcolor}
 \usepackage{wrapfig}
 \usepackage{hyperref}
 \usepackage{setspace}
 \usepackage{ulem}
 \usepackage{vmargin}
 \usepackage{tabularx}
 
 \usepackage {boxedminipage}

 \usepackage{amssymb}
 \usepackage{pifont}
 
 \usepackage{pgfplots} 
 
 \usepackage{amsthm}
 \usepackage{amsmath}
 \usepackage{amssymb}
 \usepackage{mathrsfs}
 \usepackage{lscape}
 \usepackage{slashbox}
 
  \usepackage{multirow} 
  \usepackage{booktabs}
  \usepackage{bigstrut}

 \setpapersize{A4} 
\newcommand{\xmark}{\ding{55}}%

\usepackage{listings}
\usepackage{xcolor}

\def\BibTeX{{\rm B\kern-.05em{\sc i\kern-.025em b}\kern-.08em
		T\kern-.1667em\lower.7ex\hbox{E}\kern-.125emX}}

\colorlet{punct}{red!60!black}
\definecolor{background}{HTML}{EEEEEE}
\definecolor{delim}{RGB}{20,105,176}
\colorlet{numb}{magenta!60!black}

\lstdefinelanguage{json}{
	basicstyle=\normalfont\ttfamily,
	numbers=left,
	numberstyle=\scriptsize,
	stepnumber=1,
	numbersep=8pt,
	showstringspaces=false,
	breaklines=true,
	frame=lines,
	literate=
	*{0}{{{\color{numb}0}}}{1}
	{1}{{{\color{numb}1}}}{1}
	{2}{{{\color{numb}2}}}{1}
	{3}{{{\color{numb}3}}}{1}
	{4}{{{\color{numb}4}}}{1}
	{5}{{{\color{numb}5}}}{1}
	{6}{{{\color{numb}6}}}{1}
	{7}{{{\color{numb}7}}}{1}
	{8}{{{\color{numb}8}}}{1}
	{9}{{{\color{numb}9}}}{1}
	{:}{{{\color{punct}{:}}}}{1}
	{,}{{{\color{punct}{,}}}}{1}
	{\{}{{{\color{delim}{\{}}}}{1}
	{\}}{{{\color{delim}{\}}}}}{1}
	{[}{{{\color{delim}{[}}}}{1}
	{]}{{{\color{delim}{]}}}}{1},
}

\let\chapter\section
\usepackage{amsmath}
\usepackage[linesnumbered,ruled]{algorithm2e}
\SetKwRepeat{Do}{do}{while}%
\usepackage{algpseudocode}
\algnewcommand\algorithmicforeach{\textbf{for each}}
\algdef{S}[FOR]{ForEach}[1]{\algorithmicforeach\ #1\ \algorithmicdo}

\usepackage{float} 

\date{}
 
\begin{document}
 		 \newcolumntype{M}[1]{>{\centering\arraybackslash}m{#1}}
 		 
		 \title{Storing, preprocessing and analyzing Tweets: Finding the suitable NoSQL system}
 	\author{Souad Amghar\footnote{souad.amghar@um5s.net.ma}}
 	\author{Safae Cherdal\footnote{safaecherdal@gmail.com}}
 	\author{Salma Mouline\footnote{salma.mouline@um5.ac.ma} }
 	\affil{LRIT Laboratory, Rabat IT Center, Faculty of Sciences\\ Mohammed V University in Rabat, Morocco}
 	
 	
 	
 		\maketitle
 		
 	\begin{abstract}
%

NoSQL systems are a new generation of databases that aim to handle a large volume of data. However there is a large set of NoSQL systems, each has its own characteristics. Consequently choosing the suitable NoSQL system to handle Tweets is challenging. Based on these motivations, this work is carried out to find the suitable NoSQL system to store, preprocess and analyze Tweets. 

This paper presents the requirements of managing Tweets and provides a detailed comparison of five of the most popular NoSQL systems namely, Redis, Cassandra, MongoDB, Couchbase and Neo4j regarding to these requirements. The results of this work show that for Tweets storing, preprocessing and analyzing, MongoDB and Couchbase are the most suitable NoSQL systems. Unlike related works, this work compares five NoSQL systems from different types in a real scenario which is Tweet storing, preprocessing and analyzing. The chosen scenario enables to evaluate not only the performance of read and write operations, but also other requirements related to Tweets management such as scalability, analysis tools support and analysis languages support. 
 	 	\end{abstract}


\section{Introduction}
As we live in the age of social media, a large volume of data is generated every day when we upload pictures on Instagram, post Tweets on Twitter, comment to a post on Facebook, or do any other activity on social networking sites. 
Social media data have the potential to bring smartness in many sectors, for instance, they can be used to identify management issues in food industry \cite{singh2018social}, they can also help in disaster management by propagating emergency informations \cite{kim2018social}, or be explored as a data source for public health monitoring \cite{thoma2015social}. In other words, social media data is driving big changes in different sectors and tourism is one of the most influenced  sectors. 

\vspace{0,30 cm}
In tourism, social networking platforms play an important role in travel review and recommendation \cite{memon2015travel} \cite{miguens2008social}. Furthermore,
they can be used as an alternative of traditional surveys in order to understand tourists' preferences \cite{hausmann2018social} \cite{sotiriadis2013electronic}.
Twitter is a social media platform where tourists can share their travel experiences and express their impressions about tourism destinations, hotels or restaurants using Tweets which are short messages limited in 280 characters. Consequently, Twitter plays the role of online review \cite{sotiriadis2013electronic} in tourism industry. Furthermore context-aware recommendation systems can use Tweets to determine preferences of tourists in real time \cite{meehan2013context}, this can help to recommend the suitable destination, hotel or activity. 

Tweets often contain significant informations because they are concise and, in most cases, posted in real time,  which explains the use of Tweets to perform analysis in tourism \cite{shimada2011analyzing}. 
However, the incredible amount of data captured by Twitter needs to be stored for further processing. Storing Tweets can be challenging for many database systems, because, on the one side, we have to deal with a large volume of semi-structured data, and on the other side, we have to prepare data for the analysis phase.
In this context, relational database systems that are originally created to store structured data, have many limitations in handling Tweets. 

NoSQL databases are a new generation of database systems defined in order to handle large datasets. They become a choice of many organizations that are dealing with massive data \cite{fowler2016teaching}. However we can find a large set of NoSQL systems, every one has its own approach to store and to query data. Consequently choosing the suitable NoSQL system to handle Tweets can be a challenging task. 

\vspace{0,30 cm}
In this context, our work is carried out to find the suitable NoSQL system to store, preprocess and analyze Tweets.
This paper provides a detailed comparison of five of the most popular NoSQL systems namely, Redis\footnote{redis.io}, Cassandra\footnote{cassandra.apache.org}, MongoDB\footnote{www.mongodb.com}, Couchbase\footnote{www.couchbase.com} and Neo4j\footnote{neo4j.com} regarding to Tweets management requirements. To do that, we need to answer the following questions : 
\begin{enumerate}
	\item What are the requirements of managing Tweets ?
	\item How does every NoSQL system respond to Tweets management requirements ?
\end{enumerate}
This paper is divided into seven sections. Section 2 gives a definition of NoSQL databases, as well as an overview of the five chosen NoSQL systems, namely, Redis, Cassandra, MongoDB, Couchbase and Neo4j. In section 3 we present the most important requirements of managing Tweets and we describe how the five NoSQL systems respond to these requirements. A case study is presented in section 4 which aims to evaluate the performances of the five NoSQL systems. Section 5 discusses the comparison results. And section 6 provides some related works. Finally, our conclusion is given on the last section.

\section{NoSQL Databases}
In this section, we give a definition of NoSQL databases, as well as the characteristics of the five chosen NoSQL systems, which are Redis, Cassandra, MongoDB, Couchbase and Neo4j. 

\vspace{0,30 cm}
Relational database systems have been widely used since 1970. They are based upon the ACID (Atomicity, Consistency, Isolation, Durability) model, and they use SQL (Structured Query Language) to query data, therefore they provide good performance in handling structured data. However Relational database systems  have some drawbacks in dealing with semi-structured (XML (Extensible Markup Language), JSON (JavaScript Object Notation),..) and unstructured data (documents, videos, emails,..) which represent a big part of today's data \cite{leavitt2010will}. Relational database systems have also some limits in handling huge volume of data, which requires high scalability and availability.

NoSQL database systems have been defined with the purpose to provide high scalability and availability.
The NoSQL movement is started in 2007 when Amazon published a paper about Dynamo, their key-value database system. Dynamo was created to provide high availability of Amazon's services by scarifying consistency in some cases \cite{decandia2007dynamo}.
NoSQL term (\textbf{N}ot \textbf{O}nly \textbf{SQL}) is often used to describe distributed, non-relational and scalable database systems \cite{amghar2018nosql}. NoSQL database systems are designed to handle large volume of data, therefore data are distributed across servers called nodes and they are replicated in order to insure high availability. Data replication can follow two strategies. The first strategy is the master-slave replication which consists of one master node that receives write operations, and many slave nodes that replicate the master node. The second strategy is the master-master replication where nodes play the same role.  

NoSQL databases are classified in four different categories: key-value, column-oriented, document-oriented and graph-oriented \cite{amghar2018nosql}.
There are many NoSQL database systems, each one has its own characteristics. In this work we choose five NoSQL systems, from the four categories, namely Redis, Cassandra, MongoDB, Couchbase and Neo4j based on their popularity and their large use. 

\begin{description}
	\item[Redis] is a key-value database system, created by Salvatore Sanfilippo in 2009 \cite{pokorny2013nosql}. Redis is in-memory database which means that data are loaded into memory, and they are saved  periodically to the hard disk \cite{han2011survey}. By being in-memory database, this NoSQL system provides important performances in read and write operations, however it is limited by the physical memory \cite{han2011survey}. It has a master-slave replication model and it provides SET and GET commands to respectively insert and query data \cite{Redisdb}.
	\item[Cassandra] is created by Facebook in 2008 to handle Inbox search \cite{lakshman2010cassandra}. It is a column-oriented database, which means that data are stored as rows and columns. Cassandra is no longer schema-less, which means that a table must be defined before inserting data \cite{Cassandraschema}.
	It provides a ring architecture, and data are replicated with a master-less strategy, which means that all nodes play the same role \cite{amghar2018nosql}. This NoSQL system uses its own query language, which is a SQL-like language called CQL \cite{CassandraCQL}.	
	\item[MongoDB] is created by DoubleClick in 2007 \cite{amghar2018nosql}. It is a document-oriented database systems, which means that data are stored in documents. The format in which documents are stored is  BSON which is a binary-encoded serialization of JSON. For this reason, MongoDB is a schema-less database system, this means that it is not necessary to define the database schema at the beginning.   
	Data are replicated with a master-slave model \cite{mongodoc}. MongoDB provides many methods to insert, query, update and delete documents. 
	\item[Couchbase] is created by Memcached in 2011 \cite{amghar2018nosql}. It is a document-oriented and key-value database. Each record consists of a key and a value. A value can be either binary or JSON. Binary value can only be retrieved by key, however JSON value can be parsed, indexed, and queried \cite{couchbasedoc}. In Couchbase, data are replicated with a master-master model. N1QL is the query language provided by Couchbase to query documents, it has a SQL-like syntax and semantic \cite{couchbasedoc}.	
	\item[Neo4j] is a graph-oriented database, created by Neo Technology, Inc in 2007 to handle highly connected data \cite{amghar2018nosql}. Since it is a graph-oriented database, Neo4j stores data using nodes and relations. In Neo4j, data are not distributed because nodes are highly interconnected. However, the all graph can be replicated with a master-slave replication model. Neo4j provides Gremlin and Cypher as query languages in order to create and query graphs \cite{miller2013graph}. 
\end{description}

Table \ref{tabNoSQLDB} summarizes characteristics corresponding to the five databases previously presented. 

\begin{table}[H]
	\centering
	\begin{tabular}{|M{2cm}|M{2cm}|M{2cm}|M{2cm}|M{2cm}|}
		\hline
		\textbf{Databases} &\textbf{Created by}& \textbf{Data model}& \textbf{Replication} & \textbf{Query Language} \\ \hline \hline
		\textbf{Redis} &Salvatore Sanfilippo & Key-value  & Master-Slave & GET and SET commands \\ \hline
		\textbf{Cassandra} &Facebook&Table of columns and rows&Master-Master& CQL \\ \hline
		\textbf{MongoDB} &DoubleClick &BSON Documents& Master-Slave& MongoDB methods\\ \hline
		\textbf{Couchbase} &Memcached&Key-value and JSON document&Master-Master&N1QL \\ \hline
		\textbf{Neo4j} &Neo Technology, Inc& Graph&Master-Slave&Gremlin and Cypher \\ \hline

	\end{tabular}
	\caption{NoSQL systems characteristics}
	\label{tabNoSQLDB}
\end{table} 

As we present in the current section, we choose to use different types of NoSQL systems in order to evaluate and compare all categories. However, we need first to define the requirements on which we based our comparison. In the following section, we give the most important requirements of managing Tweets and we present how every NoSQL system responds to these requirements.

\section{Tweets management Requirements}
The goal of our work is to find the most suitable NoSQL database system to store, preprocess and analyze Tweets. However, we need first to define the Tweets management requirements. For this end, we discuss in this section two requirements of storing and managing Tweets, which are scalability in addition to analysis tools and languages support, however we discuss the third requirement, namely the performance, in a detailed case study in section 4.    	
\subsection{Scalability}
Several applications use a huge volume of Tweets. Thus, database systems have to deal with growing data. 
In the context of databases, the term scalability can be defined as the capability of a database system to handle growing data. In other words, we consider a database highly scalable, when it is able to add new data without loosing its performances. This includes the capability to redistribute data when adding new node while keeping data available. 
There are two approaches to scale a database: 
\begin{itemize}
	\item[$\bullet$] Vertical scalability: called also scale-up \cite{pokorny2013nosql}, consists of adding new resources to the node, such as adding RAM (Random Access Memory), making disk input and output go faster, or by moving data to a new expensive big server, which can decrease system's performances \cite{pokorny2013nosql}. Centralized databases scale vertically. 
	\item[$\bullet$] Horizontal scalability: called also scale-out \cite{pokorny2013nosql}, consists of adding new nodes. Distributed databases, namely NoSQL databases, are designed for horizontal scalability, because they distribute data across many small machines. This makes the scalability much more easier and less expensive \cite{cattell2011scalable}. 	
\end{itemize}
To sum up, in order to handle Tweets, the database system has to be highly scalable. Since we are interested by NoSQL databases, we compare the five NoSQL systems regarding to horizontal scalability. 

\vspace{0,5 cm}
\textbf{NoSQL systems and Horizontal Scalability} \label{NoSQLAndScalability} 

\vspace{0,25 cm}
The scalability is one of the major factors that make NoSQL databases popular. However the scalability performance changes from a database system to another. 
One of the reasons that make the scalability performance higher is the database architecture. According to \cite{tang2016performance}, the Master-Master architecture systems are highly scalable, because nodes are identical, which makes the scalability less complicated.
This means, as we presented in our previous work  \cite{amghar2018nosql}, that Cassandra and Couchbase have high performance in scalability. 
The scalability is also influenced by the data model of the database. In \cite{chandra2015base}, the authors considered that graph databases, such as Neo4j, provide low scalability performance compared to other data model types. This can be explained by the fact that a graph can not be distributed across nodes, which makes the scalability more complicated. 
Another scalability factor is data auto-sharding. Auto-sharding is the capacity of a database system to distribute data across nodes automatically. In Redis, data have to be distributed by the user \cite{corbellini2017persisting}, which makes Redis scalability poor \cite{han2011survey}. However, Cassandra \cite{kan2014cassandra}, Couchbase \cite{zahid2014security} and MongoDB \cite{gu2015application} provide  automatic data distribution across nodes. 
To sum up, Couchbase and Cassandra provide better performance in scalability compared to the other systems, MongoDB provides important performance too, however Neo4j and Redis have some limits in scalability. 

Table \ref{tabNoSQLDBscalability} summarizes the  comparison between the five NoSQL databases relatively to scalability
\begin{table}[H]
	\centering
	\begin{tabular}{|M{2cm}|M{1.9cm}|M{1.9cm}|M{1.9cm}|M{1.9cm}|M{1.9cm}|}
		\hline
		Databases &\textbf{Redis}&\textbf{Cassandra}&  \textbf{MongoDB}& \textbf{Couchbase}&\textbf{Neo4j} \\ \hline 
		Scalability & - & ++  & + & ++ & - \\ \hline	
	\end{tabular}
	\caption{NoSQL systems \& Horizontal Scalability}
	\label{tabNoSQLDBscalability}
\end{table} 

\subsection{Analysis tools and languages support }
The main goal of gathering Tweets is to analyze them. For this reason, database systems have to support analysis tools connectivity and to provide modules and libraries for analysis languages.
There is a lot of analysis tools such as Hadoop\cite{taylor2010overview}, Apache Spark \cite{shoro2015big}, and Apache storm \cite{iqbal2015big}: 
\begin{itemize}
	\item [$\bullet$]Hadoop is a software framework that provides large scale distributed data analysis \cite{taylor2010overview}. Hadoop provides HDFS (Hadoop Distributed File System ), which is a master-slave architecture that stores data and executes read and write instructions. Nevertheless, in some applications, we need to use other database systems instead of, or with, HDFS. 
	\item [$\bullet$]Apache Spark is a unified engine for distributed data processing \cite{zaharia2016apache} which makes data analysis faster \cite{shoro2015big}. It provides API (Application Programing Interfaces) in many programming languages and also supports many tools including structured data processing (Spark SQL), machine learning (MLlib), graph processing (GraphX), and Spark Streaming \cite{Spark}.
	\item [$\bullet$]Apache Storm is a stream processing system that can process unbounded streams of data very fast \cite{iqbal2015big}. 
	Storm applications are called topologies. A Storm topology is a graph of tasks that process distributed streams of data. As long as it is not stopped by the user, a topology can run forever \cite{peng2015r}.
\end{itemize}
The most used analysis languages are R language \cite{Rproject} and Python language \cite{sanner1999python}:
\begin{itemize}	
	\item [$\bullet$]R is an analysis language for statistical computing and graphics, it is similar to the S language \cite{chambers1998programming}. R provides statistical and graphical techniques and  it allows users to add more functionality by defining new functions \cite{Rproject}.
	\item [$\bullet$]Python is one of the most used languages for analysis thanks to its object oriented design and modularity \cite{sanner1999python}, it is both simple and powerful. This makes it easier to work with Python in many contexts. 
	
\end{itemize}

\vspace{0,25 cm}
\textbf{NoSQL systems \& Analysis tools and languages} 

\vspace{0,25 cm}
The ability to connect to analysis tools and languages makes the database system more suitable to store, manage and analyze Tweets. For this reason we discuss in the following the drivers, connectors and libraries provided to ensure this integration.

\begin{itemize}
	\item [$\bullet$] Redis can be used with Hadoop as Cache Server \cite{scholarperformance}, which can improve the performance of MapReduce job, since accessing data from Cache is faster than disk, it is used in several works \cite{mattmann2017scalable} \cite{scholarperformance}.
	Redis provides also the spark-redis package, which is the Redis connector for Apache Spark \cite{Redisdb}. Spark Redis integration is used in many works such as \cite{aydin2017batch} where authors use Redis and Spark to collect and analyze real time data. 
	Storm adopter can use Storm-redis  to either retrieves or store values into Redis \cite{storm}. For instance, Redis and Storm are used in \cite{song2017model} to create a framework of data processing applications. Moreover, Redis can be used with R language using Rredis package \cite{Rproject} and it can be used with Python using redis-py module \cite{lewis2010key}. 
	
	\item [$\bullet$] Cassandra can be used with Hadoop as the input or output of map reduce jobs by doing some setups and configurations \cite{capriolo2011cassandra}. \cite{dede2016processing} and \cite{vasavi2018framework} are examples of works that use Cassandra and Hadoop integration .
	To connect Spark to Cassandra, we need to add Cassandra connector to the Spark project \cite{chaudhari2019scsi}, this integration can be used  to benefit from the consistency storage of Cassandra and the fast processing of Spark \cite{llados2018scalable}. In other hand, Storm-cassandra is a library that enable to work with Storm on top of Cassandra \cite{storm}, this integration is used in \cite{huang2017big} to define a  data processing platform for intelligent agriculture. 
	Finally, to connect Cassandra with R language, we can use Rcassandra which is a package provided by R-project \cite{Rproject}, however, Python programmers can use Cassandra-Driver module to work with Cassandra \cite{Rproject}. 
	
	\item [$\bullet$] MongoDB and Hadoop integration has been used in many applications for scientific data analysis \cite{dede2013performance}, Time-dependent shortest path calculation \cite{boulmakoul2014mongodb}, and other application fields \cite{dai2013design}. However, this integration is no longer supported by MongoDB \cite{mongodoc}. 
	MongoDB provides a connector for Spark to use all Spark libraries \cite{mongodoc}. MongoDB and Spark are used in \cite{le2016efficient} to analyze data collected from smart clothes, this integration is also used in \cite{ocampo2018scalable} to define an architecture for traffic monitoring.
	Storm provides a package that enables to work with MongoDB, it enables to both insert and update data \cite{storm}. Storm and MongoDB are used in a real-time framework to analyze self-health data collected from wearable devices \cite{cai2015real}. Finally, RmongoDB is a MongoDB client for R, which is used in \cite{hassan2017sentimental} to analyze comment on e-commerce websites, and Pymongo is a module for Python which is used in variety of works, among them Tweets analysis \cite{nayak2014mongodb}.
	
	\item [$\bullet$] Couchbase provides Hadoop Connector to connect Hadoop with the 2.5, 3.x and 4.x versions of Couchbase.   
	However, Hadoop is no longer supported for the 5.x and 6.x versions of Couchbase, because Hadoop uses the TAP feed API which is removed from the last version of Couchbase  \cite{couchbasedoc}. 
	Couchbase provides a Spark connector to work with Apache Spark \cite{couchbasedoc}, this integration is used in \cite{asri2017real} to predict  miscarriages in real time. However, neither Couchbase nor Apache Storm provides a Couchbase Storm integration. Similarly to Storm, there is no package to use Couchbase with R, however we can use Couchbase with Python by using the Couchbase Python Client \cite{kagramanyan2017document}.
	
	\item [$\bullet$] Neo4j and Hadoop integration is used in many works, among them web crawling \cite{pratiba2017distributed} and power network topology processing \cite{lv2017storage}. This is the suitable choice when working with networked data.
	Neo4j also provides Neo4j-Spark-Connector that enables to transfer data from and to a Neo4j database \cite{neodoc}. Neo4j and Spark are used in \cite{do2018w} to evaluate the similarity between objects. 
	Similarly to Couchbase, there is no integration of Neo4j and Storm. However, Neo4r is a Neo4j driver that allows to use Neo4j with R \cite{Rproject}, and Neo4j-driver is a Python module that can be used with Python. Time-varying Social Networks \cite{cattuto2013time} and emotion analysis \cite{drakopoulos2016tensor} are examples of works that implements Python with Neo4j.
	
\end{itemize}
Table \ref{tabNoSQLDBAnalysisTools} is a summary of NoSQL systems analysis tools support.
\begin{table}[H]
	\centering
	\begin{tabular}{|M{1.8cm}|M{1.7cm}|M{1.9cm}|M{1.9cm}|M{2.5cm}|M{1.7cm}|}
		\hline
		Databases &\textbf{Redis}&\textbf{Cassandra}& \textbf{MongoDB}& \textbf{Couchbase}&\textbf{Neo4j} \\ \hline 
		Hadoop & \checkmark & \checkmark  & \xmark & \xmark (5.0+) ; \checkmark (2.5, 3.x 4.x) & \checkmark  \\ \hline
		Spark & \checkmark &\checkmark &\checkmark &\checkmark &\checkmark \\ \hline
		Storm &\checkmark &\checkmark &\checkmark & \xmark & \xmark \\ \hline
		R language &\checkmark &\checkmark &\checkmark &\xmark &\checkmark \\ \hline
		Python language &\checkmark& \checkmark &  \checkmark &  \checkmark & \checkmark  \\ \hline
		
	\end{tabular}
	\caption{NoSQL databases \& Analysis tools and languages}
	\label{tabNoSQLDBAnalysisTools}
\end{table} 

In this section we discussed two Tweets management requirements, which are scalability in addition to analysis tools and languages support.  In order to evaluate the performance of the five NoSQL systems in details, we present in the following section a case study of collecting, storing, preprocessing and analyzing Tweets.

	\section{NoSQL systems and Performance: A case study}
Analysis applications need to provide good performance in term of execution time. One of the factors that influence the application performance is the database \textit{read} and \textit{write} performances. NoSQL database systems provide, relatively, high performance of \textit{read} and \textit{write} operations, but, there is a difference from a database system to another. 
\subsection{Presentation of the case study}

In order to compare the five NoSQL systems' performances, we conduct an experiment where we use them in storing, preprocessing and analyzing Tweets. This experiment aims to evaluate \textit{write} and \textit{read} performances of the five NoSQL systems through a case study that aims to analyze Tweets about tourism in Morocco. 
To this end, we first collect Tweets containing the most visited cities in Morocco, then we insert the collected Tweets in the five NoSQL systems which enables to evaluate \textit{write} performance of each system. After that, a preprocessing phase is needed to prepare data to be finally analyzed. Tweets preprocessing and analyzing enable to evaluate \textit{read} and \textit{write} performances of the used systems. 

In this case study, Tweets are processed using a Python programming interface. This choice is based on the language simplicity and the ability of use with the five NoSQL systems. Moreover, the interfaces used in Tweets insertion, preprocessing and analyzing are the same for the five NoSQL systems. As a consequence, the database system is what makes the difference. 

All experiments are executed in one machine with the following configuration:
\begin{itemize}
	\item [$\bullet$]CPU: Intel(R) Core(TM) i7-7700 CPU @ 3.60 GHZ
	\item [$\bullet$]RAM: 16 GB
	\item [$\bullet$]Operating system: Ubuntu 18.04.1 LTS
\end{itemize}
Table \ref{tabNoSQLDBPython2} summarizes the database systems and Python module used. 
\begin{table}[H]
	\centering
	\begin{tabular}{|M{2cm}|M{1.9cm}|M{1.9cm}|M{1.9cm}|M{1.9cm}|M{1.9cm}|}
		\hline
		Databases &\textbf{Redis}  &\textbf{Cassandra}&  \textbf{MongoDB}& \textbf{Couchbase}& \textbf{Neo4j}\\ \hline 
		Database version & 5.0.3 & 3.11.3 &  4.0.5 &5.0.1 & 3.5.2 \\ \hline
		Python module &redis-py  & Cassandra-Driver & Pymongo  &Couchbase Python Client & neo4j-driver \\ \hline

	\end{tabular}
	\caption{NoSQL databases Python Modules}
	\label{tabNoSQLDBPython2}
\end{table}

\subsection{Tweets collection}
Twitter provides an API to harvest Tweets on a specific topic. By using a programming language and Twitter API, we can collect a large volume of Tweets to match with a specific topic. Tweets are mainly generated in JSON format, however, we can use any other format such as CSV (Comma Separated Values).  

In this work, we are interested by the tourism of Morocco. Hence, we aim to collect Tweets containing Moroccan cities. To this purpose, 
we defined a Python programming interface to collect Tweets using a list of Moroccan cities (Marakech, Agadir, Casablanca, Chefchaouen...). As a response, a JSON file is generated containing more than 1000000 Tweets \footnote{You can find the used Tweets at https://sites.google.com/view/souadamghar/publications?authuser=0}. Additionally to the text, each Tweet contains several properties such as Tweet ID, creation date, user ID, user name, and user description. 

Listing  \ref{l1} is an example of the collected Tweets.

\begin{lstlisting}[language=json,firstnumber=1,caption={Example of a collected Tweet},captionpos=b, label={l1}]
{'created_at':'Thu May 24 17:03:40 +0000 2018',
'id':999000850006580020,
'id_str':'999000850006580020',
'text':'@Corderagirl hope you having a good time in Rabat',
'display_text_range':[13,125],
'source':'\u003ca href=\'http:\/\/twitter.com\/download\/iphone\' rel=\'nofollow\'\u003eTwitter for iPhone\u003c\/a\u003e',
'user':{
'id':777772036841289728,
'id_str':'777772036841289728',
'name':'National Hunt Pete',
'screen_name':'BumperToJumper',
'location':'The world is my Playground'
}
...
}
\end{lstlisting}

\subsection{Write performances for Tweets insertion}
In this experiment, we need to insert data from a JSON file in the five databases. Tweets insertion enables to compare the write performances of the five NoSQL systems. However, not all these NoSQL systems provide tools that enable to load  a JSON file to the database.
MongoDB provides \textit{mongoimport} tool which imports data from  Extended JSON, CSV, or TSV (Tabulation-Separated Values) files. 
It also provides an exporting tool called \textit{mongoexport} that exports mongodb documents to a JSON, CSV or TSV files.
These two tools are very useful in data management. 
Couchbase provides \textit{cbdocloader} command to load data from JSON file, however it loads a group of JSON documents in a given directory or in a single zip file into a Couchbase bucket, which means that every JSON file has to contain only one document which is not the case with our data. Which means that we need to transform the JSON file before using \textit{cbdocloader}. 
Cassandra, Neo4j and Redis do not provide any tool to import data from a JSON file. For this reason we use, for every database system, an interface that loads data from JSON file and inserts them into the database. 

\vspace{0,30 cm}
As we mentioned previously, the JSON file generated by Twitter API contains a lot of informations about the Tweet and the user who posted it. Furthermore, the number of informations can change from a Tweet to another. For instance, if the Tweet is retweeted, which means that is forwarded, we will find 34 additional informations about the retweet and the user who reposted it. Thus, a unified schema is needed, not only because there is a plenty of informations that we do not need in the analysis, but also because not all the five NoSQL systems can handle a dynamic schema dataset. Consequently, in order to have the same schema for all systems, we keep only the following fields: Tweet identifier (id\_str) , date of Tweet (created\_at), Tweet message (text), user identifier (user.id\_str), user name (user.name), user description (user.description), user location (user.location). 

As a result, the database models are as follow:
\begin{itemize}
	\item In Redis, Tweets' informations are stored in a hash that stores a map of keys and values as shown in Fig \ref{redisSchema}. 
	\begin{figure}[H]
		\centering
		\includegraphics[width=0.5\textwidth]{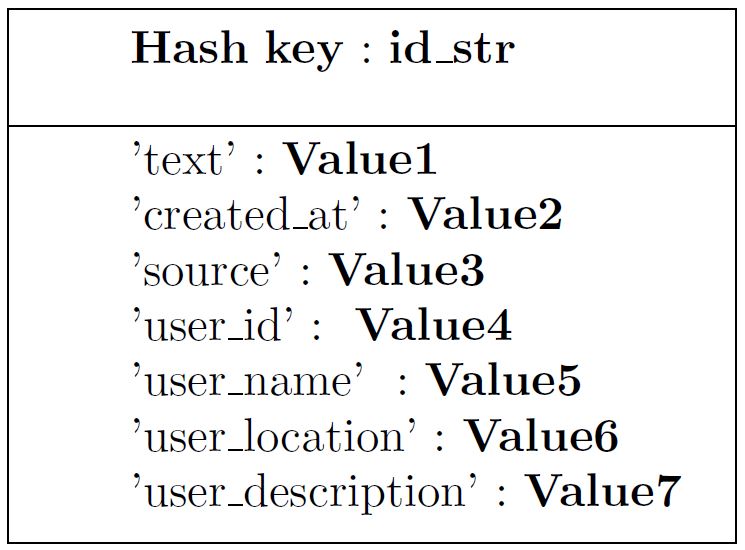}
		\caption{Redis Schema} \label{redisSchema}
		
	\end{figure}
	
	\item In Cassandra we stored the data in a table called \textit{Tweets} containing eight columns, as follow:
	
	\textit{\textbf{Tweets( tweet\_id, created\_at, tweet\_text, tweet\_source, user\_id, user\_name ,  user\_location ,  user\_description )}}. Where \textit{tweet\_id} is the primary key.
	\item In MongoDB we store data in a collection called Tweets which contains data records in Binary JSON format, as shown in Fig \ref{MCSchema}  .
	\item In Couchbase, similarly to MongoDB, Tweets informations are stored in JSON format in a bucket called Tweets, as shown in Fig \ref{MCSchema}. 
	\begin{figure}[H]
		\centering
		\includegraphics[width=0.5\textwidth]{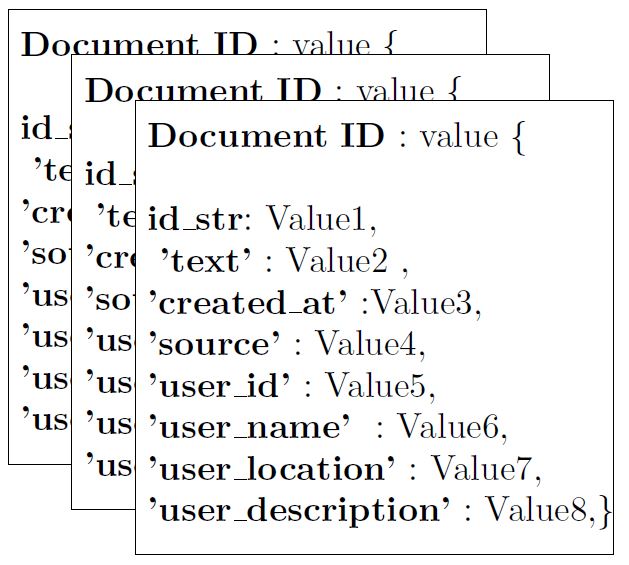}
		\caption{MongoDB and Couchbase Schema} \label{MCSchema}
	\end{figure}
	
	\item  In neo4j we stored the data on two Nodes, the first node is called \textit{Tweet} and it contains informations about the Tweet. The second node is called \textit{User} and it contains the user informations as shown in the Fig \ref{neoSchema}.
	\begin{figure}[H]
		\centering
		\includegraphics[width=0.8\textwidth]{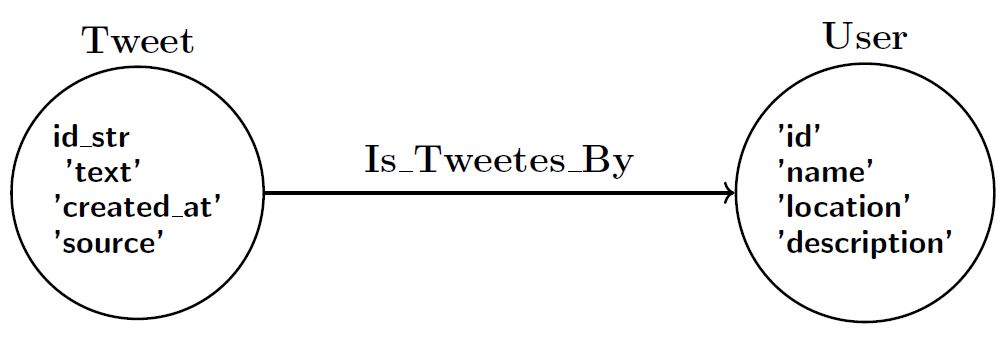}
		\caption{Neo4j Schema} \label{neoSchema}
	\end{figure}
	
\end{itemize}

To insert Tweets into the five NoSQL systems, we have written a python script that extracts Tweet by Tweet from the JSON file and insert the selected fields into the database as shown in the algorithm \ref{algoInsertion}. 

\begin{algorithm}
	\SetKwInOut{Input}{Input}
	\SetKwInOut{Output}{Output}
	
	\underline{function InsertTweets} $(JSONFile)$\;
	\Input{A JSON File}
	
	\ForEach{Line in  JSONFile}{ 
		Insert Line into Database\;
	}
	
	\caption{Insert Tweets into NoSQL Database from a JSON file} \label{algoInsertion}
\end{algorithm}

For every NoSQL system, we use a method defined by the Python Driver in order to insert data into the database. 
For Redis, we use the \textit{HSET} command which enables to store every Tweet in a hash with a unique key which is the Tweet id (id\_str). The other Tweets informations (text, source..) are added to the hash as key-value pairs.
For Cassandra, we use the \textit{execute} method that, as its name implies, executes CQL commands such as \textit{INSERT INTO} used in our case.  With MongoDB, we use the method \textit{insert\_one} that enables to insert a JSON document into the database. 
Similarly to MongoDB, we use the \textit{insert} method in order to insert a JSON document into Couchbase.
For Neo4j, we use the \textit{run} method that aims to execute a Cypher command which creates the graph and inserts data into it.  

We execute the code using different numbers of Tweets in order to compare the five databases relatively to different sizes of data. Table \ref{dbsInsertion} summarizes the insertion time in seconds. \\

\begin{table}[H]
	\centering
	\begin{tabular}{|M{2cm}|M{1.9cm}|M{1.9cm}|M{1.9cm}|M{1.9cm}|M{1.9cm}|}
		\hline
		Number Of Tweets &\textbf{Redis}&\textbf{Cassandra}&\textbf{MongoDB} &\textbf{Couchbase} &\textbf{Neo4j} \\ \hline 
		10000 & 2.75 &4.60&3.06&\textbf{1.54}& \textbf{175.16}\\ \hline
		100000 &27.15 &36.98 &28.59 &\textbf{15.31}&\textbf{1760}\\ \hline
		500000 &135.47 &181.55& 140.93 &\textbf{76.06}&\textbf{7043}\\ \hline
		1000000 & 287.38&398.94& 453.27 &\textbf{142.08}&\textbf{13854}\\ \hline
	\end{tabular}
	\caption{Time for Tweets insertion in the five NoSQL systems (in seconds)}
	\label{dbsInsertion}
\end{table}

The results of Tweets insertion reveal that Couchbase takes the lowest time for Tweets insertion. The time that Tweets insertion takes for Couchbase is approximately half the time that it takes for Redis, Cassandra and MongoDB. However, Neo4j is significantly slower than the other systems. 

\vspace{0,3 cm}
MongoDB and Couchbase are the only database systems, from the chosen systems, that use JSON as data model, and both of them have commands that help to load data from JSON files automatically. For this reasons, we compare the write performances of MongoDB and Couchbase using respectively \textit{mongoimport} and \textit{cbdocloader} tools. furthermore, in this experiment, we use all fields of Tweets.
Table \ref{MCInsertion}, shows that MongoDB has better \textit{write} performance when using \textit{mongoimport} tool. Moreover, to insert Tweets using \textit{cbdocloader} tool, we need to divide the single file that contains 1000000 Tweets into 1000000 files, each contain one document. 

\begin{table}[H]
	\centering
	\begin{tabular}{|M{2cm}|M{1.5cm}|M{1.5cm}|M{1.5cm}|M{1.5cm}|M{1.5cm}|M{1.5cm}|}
		\hline
		Databases &  10000 Tweets& 100000 Tweets& 500000 Tweets&1000000 Tweets \\ \hline 
		MongoDB  &\textbf{1.00} &\textbf{9.00} &\textbf{158.00} &\textbf{319.00} \\ \hline
		Couchbase &4.00 &35.00 & 521.00 &2540.00 \\\hline
	\end{tabular}
	\caption{MongoDB and Couchbase insertion time (in seconds) using \textit{mongoimport} and \textit{cbdocloader} tools}
	\label{MCInsertion}
\end{table}

After inserting Tweets in the five NoSQL systems and compare the time that it takes for every system, we present in the following Tweets preprocessing that helps to clean the Tweets' text before being analyzed.

\subsection{Read and write performances for Tweets preprocessing}

Tweet preprocessing is an important step that prepares the Tweets for the further processing. In this step we remove spatial characters, uppercase, emojis  in order to make the text readable by the analysis algorithm.
In our case study, we create a python code that reads the text from the database, cleans the text using \textit{tweet-preprocessor} Python module, and stores the new text in a new field called \textit{cleanedText} as described in the algorithm \ref{algopreprocessing}. Consequently, Tweets preprocessing enables to compare \textit{read} and \textit{write} performances of the five NoSQL systems. Both in Tweets prepocessing and analyzing, we have used datasets of 1000000 Tweets. 

For Schema-less databases namely MongoDB, Couchbase and Redis, we create the  new field (\textit{cleanedText}) during the preprocessing phase. However, for Cassandra and Neo4j we have to create this field when we define the database model, we modify then its value by inserting the new text. 

\begin{algorithm}
	\SetKwInOut{Input}{Input}
	\SetKwInOut{Output}{Output}
	
	\underline{function PreprocessTweets} $()$
	
	query $\gets $ Select Tweets' text from the database
	
	\ForEach {text in query}
	{cleanedText $\gets $ clean(text) \;
		Insert CleanedText into the database\;
	}
	
	\caption{Preprocessing Tweets} \label{algopreprocessing}
\end{algorithm}

Different queries are used to \textit{read} Tweets' text from the five databases and to insert the cleaned texts into the databases. 
For Redis, we first use the command \textit{keys} that read all the keys of the database. Then we use these keys in the command \textit{hget} in order to read the value associated with the field \textit{text}, and we use the command \textit{hset} to insert the cleaned text into the database. 
For Cassandra, we use the CQL command \textit{SELECT .. FROM} to select Tweets' text and we use the command \textit{UPDATE.. SET ..} to insert the cleaned text into the database. 
Regarding MongoDB, we use the method \textit{find()} that selects documents from the collection and we use the method \textit{update()} to modify the document by inserting the cleaned text. 
In Couchbase, we use the \textit{N1QL} query \textit{SELECT .. FROM} to select Tweets' text, and we use the method \textit{mutate\_in} that enables to modify documents in order to insert the cleaned text. 
For Neo4j, we use the Cypher command \textit{MATCH .. RETURN} to select Tweets' text and we use the command \textit{SET} to insert the cleaned Text. 

\vspace{0,3 cm}
Table \ref{dbscleaning} presents the time in seconds that Tweets preprocessing takes using each of the five databases. \\

\begin{table}[H]
	\centering
	\begin{tabular}{|M{2cm}|M{1.9cm}|M{1.9cm}|M{1.9cm}|M{1.9cm}|M{1.9cm}|}
		\hline
		Databases &\textbf{Redis}&\textbf{Cassandra}&\textbf{MongoDB} &\textbf{Couchbase} &\textbf{Neo4j} \\ \hline 
		text preprocessing &\textbf{96.45}& 314.02&314.94&119.96& 46450 \\ \hline
	\end{tabular}
	\caption{Time for Tweets Preprocessing (in seconds) }
	\label{dbscleaning}
\end{table}

Redis followed by Couchbase are the faster systems in Tweets preprocessing. MongoDB and Cassandra have proximately the same performances. However Neo4j has the lowest performance in Tweet preprocessing. 

\subsection{Read performances for Tweets analysis}
In order to compare \textit{read} performances of the five NoSQL systems, we have analyzed the collected Tweets using a sentiment analysis. 
Sentiment analysis enables to identify the polarity of a text, by classifying it into positive, negative and neutral.
The interface we create for sentiment analysis enables to read Tweets text from the database and apply sentiment analysis as shown in the algorithm \ref{algosentiment}.

\begin{algorithm}
	\SetKwInOut{Input}{Input}
	\SetKwInOut{Output}{Output}
	
	\underline{function SentimentAnalyzeTweets} $()$\;
	
	\For{ \textbf{all} CleanedText in the Database}
	{
		apply sentiment analysis \;
	}
	show results in plot \;
	
	\caption{ Tweets sentiment Analysis}\label{algosentiment}
\end{algorithm}

We have implemented the sentiment analysis using NLTK (Natural Language Toolkit), which returns for each Tweet the probabilities of it being negative, neutral and positive. The tool uses two classifiers, a Naive Bayes Classifier and a Hierarchical Classifier \cite{pletea2014security}.

Table \ref{dbsanalysis} represents the time that it takes to execute the  sentiment analysis using the five databases.

\begin{table}[H]
	\centering
	\begin{tabular}{|M{2cm}|M{1.9cm}|M{1.9cm}|M{1.9cm}|M{1.9cm}|M{1.9cm}|}
		\hline
		Databases &\textbf{Redis}&\textbf{Cassandra}&\textbf{MongoDB}  &\textbf{Couchbase} &\textbf{Neo4j} \\ \hline 
		Sentiment Analysis &803.97 &\textbf{742.58}&743.06&748.64&811.28 \\ \hline
		
	\end{tabular}
	\caption{Time for Tweets sentiment analysis (in seconds)}
	\label{dbsanalysis}
\end{table}
The obtained results of Tweets sentiment analysis show that Cassandra and MongoDB followed by Couchbase are faster compared to the other systems. Neo4j and Redis have reasonable performances. 
\section{Discussion}
The goal of the current work is to find the appropriate NoSQL system regarding to Tweets management requirements, which are scalability, analysis tools support, analysis languages support, and performance.\\
After comparing the five chosen NoSQL systems in the previous sections, we summarize in the following the comparison results for each system:

\begin{description}
	\item[Redis] has important performances in all test especially Tweets preprocessing. This can be explained by the fact that Redis loads all data in memory, which can be at the same time a drawback, because it limits the volume of data that can be stored in Redis. As presented in subsection \ref{NoSQLAndScalability}, Redis has also some limits regarding to scalability. However, in term of analysis tools and languages support, Redis can operate with all tools and languages previously presented.
	
	\item[Cassandra] is quite efficient with Tweets analyzing which means that it has important performances in data reading as Tweets analyzing is based on data reading. Cassandra has also significant performances in scalability and it can work with all analysis languages and tools reported in this paper. The only limit of Cassandra is that we need to define the database schema before inserting data which can cause some drawbacks in handling semi-structured data. 
	
	\item [MongoDB] has also significant performances in Tweets insertion by using the \textit{mongoimport} command, and Tweet analyzing as well. Another advantage of MongoDB is being a document-oriented system which enables to handle Tweets more efficiently. In term of scalability, MongoDB has important performances, and it can integrate the most of analysis tools and languages except Hadoop which is no longer supported by MongoDB.
	
	\item[Couchbase] is another document-oriented system, which means that is quite efficient in storing JSON data such as Tweets. it has also important performances in Tweets insertion as well as scalability. Couchbase provides \textit{cbdocloader} tool in order to import data from JSON files automatically, however we need to prepare a JSON file before using the tool. Moreover Couchbase can not integrate all analysis tools and languages namely Storm and R language. 
	
	\item [Neo4j] has the lowest performances in all tests, especially in Tweets insertion and preprocessing where the difference between Neo4j and the other systems is huge. Neo4j has also some drawbacks in scalability. However it can work with all analysis tools and languages that we presented in this work, except Storm. 
	
\end{description}

Based on these results, we can conclude that MongoDB and Couchbase are the most suitable in storing, preprocessing, and analyzing Tweets.

\section{Related Works}
There is considerable existing works that compare NoSQL systems. In this section, we present some related works that compare NoSQL systems characteristics. 

In \cite{chang2018sql}, authors compare between Mysql\footnote{www.mysql.com/fr/} as a SQL database system and MongoDB as a NoSQL database system in term of performance. In this comparison, Tweets are used as a dataset to compare the performance of input, retrieve and delete operations of each database. The results of this comparison show that MongoDB performance is better than MySQL in all operations. This research compares performances of SQL and NoSQL databases in storing Tweets, however, it has tended to focus on one NoSQL system. 

Another comparative analysis of NoSQL database systems is provided in \cite{ul2018performance}. In this work, six NoSQL database systems namely, AeroSpike\footnote{www.aerospike.com}, BerkeleyDB\footnote{www.oracle.com/database/technologies/related/berkeleydb.html}, CouchBase, HBase \footnote{hbase.apache.org}, MongoDB and Redis, are compared regarding to metrics such as performance, Memory consumption, ease of use, nature of data and integration of processing platforms. The authors use a dataset of unstructured medical data, containing medical tests, credit card data, office documents (word, excel), video and audio streams. The results of this comparison show that MongoDB has the highest use of memory (89\%), while Redis has the lowest use of memory (18\%). For loading data from the database to RAM, AeroSpike has the best performance among all database systems, however Berkeley has the best performance in data querying. This comparison seems to be useful because a large dataset is used. However they do not use all NoSQL systems categories. Another difference between this comparison and our work is data type. 

In \cite{pereira2018nosql}, Pereira and colleagues compare three document-oriented NoSQL systems: Couchbase, MongoDB and RethinkDB\footnote{www.rethinkdb.com}. In their performance evaluation experiments, authors compare \textit{write}, \textit{read}, \textit{update} and \textit{delete} performances of the three chosen systems. These operations are evaluated in two scenarios. One thread scenario, where one request is sent to the database in order to evaluate the response time of the three systems. However, the multiple threads scenario aims to verify how Couchbase, MongoDB and RethinkDB work with multiple requests in the same time. 
The results of this comparison show that Couchbase has a better performance at the majority of the operations, however, MongoDB scores better just in writing and reading documents with multiple threads. The goal of this comparison is to evaluate how these NoSQL systems works in real time, but it focuses on document-oriented systems while there are other types of NoSQL systems that seem to have important performances in real time. 

\vspace{0,3 cm}
Table \ref{tabarticles} summarizes all the related works reported in the current section. 
\begin{table}[H]
	\centering
	\begin{tabular}{|M{2.2cm}|M{2.4cm}|M{2.4cm}|M{2.4cm}|M{2.4cm}|}
		\hline
		&\textbf{This work (2019)} &\textbf{\cite{chang2018sql} (2018)} &\textbf{\cite{ul2018performance}(2018)} & \textbf{\cite{pereira2018nosql} 2018} \\ \hline  \hline
		
		Database systems & 5 NoSQL databases & MongoDB and MySQL &5 NoSQL databases & 3 NoSQL databases\\ \hline 
		Dataset & Tweets & Tweets & unstructured medical data & JSON documents \\ \hline
		Nomber of records &1000000 & 1000000 & 3.4 TB & not given\\ \hline
		Evaluated caracteristics & Scalability ; Analysis tools/languages support; write and read performances  &Analysis tools support; write and read performances &	Analytic tool compatibility; read Performance; Memory consumption; Ease of use & write, read and delete operations; single thread scenario and multiple threads scenario  \\ \hline

	\end{tabular}
	\caption{Summary of related works}
	\label{tabarticles}
\end{table}

In the literature, there are many works that compare NoSQL systems from different points of view. However there is no work that compare NoSQL database systems regarding to Tweets management requirements. In this paper, we compare five NoSQL systems from different types in a real scenario which is Tweet storing, preprocessing and analyzing. The chosen scenario enables to evaluate not only the performance of \textit{read} and \textit{write} operations, but also other requirements related to Tweets management such as scalability, analysis tools support and analysis languages support.

\section{Conclusion and Futur Works}	
In this paper, we provided a detailed comparison of five NoSQL systems namely, Redis, Cassandra, MongoDB, Couchbase and Neo4j regarding to Tweets management requirements which are scalability, analysis tools support, analysis languages support, and performance. We presented a detailed discussion of the five NoSQL systems' scalability performances and we reported the analysis tools and languages supported by these systems. Moreover we conducted an experiment where we use the five NoSQL systems in storing, preprocessing and analyzing Tweets. This experiment aims to evaluate \textit{write} and \textit{read} performances of the five NoSQL systems through a case study that aimed to analyze Tweets about tourism in Morocco. The results of this work shows that MongoDB and Couchbase are the most suitable NoSQL systems to store preprocess and analyze Tweets. 

Our future works will concentrate on how to unify the access to data stored in MongoDB and Couchbase in order to be able to use both of them within a single application more easily and efficiently.
 
 \bibliography{biblioA2}
 \bibliographystyle{plain}
 \end{document}